# Identifying single influential publications in a research field:

# New analysis opportunities of the CRExplorer

Andreas Thor*, Lutz Bornmann**$, Werner Marx***, Rüdiger Mutz****

$ Corresponding author

*University of Applied Sciences for Telecommunications Leipzig
Gustav-Freytag-Str. 43-45,
04277 Leipzig, Germany.
Email: thor@hft-leipzig.de

**Division for Science and Innovation Studies
Administrative Headquarters of the Max Planck Society
Hofgartenstr. 8,
80539 Munich, Germany.
Email: bornmann@gv.mpg.de

***Max Planck Institute for Solid State Research
Information Service
Heisenbergstrasse 1,
70506 Stuttgart, Germany.
Email: w.marx@fkf.mpg.de

****ETH Zürich
Mühlegasse 21
8001 Zurich, Switzerland
Email: mutz@gess.ethz.ch

**Abstract**

Reference Publication Year Spectroscopy (RPYS) has been developed for identifying the cited references (CRs) with the greatest influence in a given paper set (mostly sets of papers on certain topics or fields). The program CRExplorer (see www.crexplorer.net) was specifically developed by Thor, Marx, Leydesdorff, and Bornmann (2016a, 2016b) for applying RPYS to publication sets downloaded from Scopus or Web of Science. In this study, we present some advanced methods which have been newly developed for CRExplorer. These methods are able to identify and characterize the CRs which have been influential across a longer period (many citing years). The new methods are demonstrated in this study using all the papers published in *Scientometrics* between 1978 and 2016. The indicators N_TOP50, N_TOP25, and N_TOP10 can be used to identify those CRs which belong to the 50%, 25%, or 10% most frequently cited publications (CRs) over many citing publication years. In the *Scientometrics* dataset, for example, Lotka's (1926) paper on the distribution of scientific productivity belongs to the top 10% publications (CRs) in 36 citing years. Furthermore, the new version of CRExplorer analyzes the impact sequence of CRs across citing years. CRs can have below average (-), average (0), or above average (+) impact in citing years (whereby average is meant in the sense of expected values). The sequence (e.g. 00++---0--00) is used by the program to identify papers with typical impact distributions. For example, CRs can have early, but not late impact ("hot papers", e.g. +++---) or vice versa ("sleeping beauties", e.g. ---0000---++).

# 1 Introduction

Research activity is usually based on previous investigations in a scientific community: "Original ideas seldom come entirely 'out of the blue'. They are typically novel combinations of existing ideas" (Ziman, 2000, p. 212). Findings are re-combined and developed further, resulting in scientific progress. According to Popper (1961), knowledge is acquired when hypotheses are formed using earlier findings and empirically tested. According to the alternative view of Kuhn (1962) hypotheses are formulated and empirically tested within paradigms or exemplars, which provide frameworks within which specific puzzles are solved (see here also Abbott, 2001). Paradigms are "a set of guiding concepts, theories and methods, on which most members of the relevant community agree" (Kaiser, 2012, p. 166). Whereas Kuhn (1962) sees scientific progress as changes of paradigms in a non-cumulative process, for Popper (1961) progress is a cumulative process. Despite the fundamental differences of the two approaches to explaining scientific progress, in principle progress is not possible in either approach without the cognitive influence on current research of past literature.

The influence of past literature on current research is manifested by references cited in publications. Thus, the premise of the normative theory of citations is that the more frequently a particular publication is cited, the more important it is for scientific progress (Bornmann, de Moya-Anegón, & Leydesdorff, 2010; Merton, 1965). This premise is not only the foundation for the use of citation counts in research evaluation (Bornmann & Daniel, 2008), but also the use of cited reference (CR) counts to analyze the historical roots of research fields and topics (Marx & Bornmann, 2016). Bornmann and Marx (2013) proposed changing the perspective of the classic times cited analysis (which is a forward view) to the perspective of major historical contributions to a specific research field (which is a backward view). In the backward view, the number of times CRs are cited in publications of a given research field is analyzed. Of course, both perspectives are closely interconnected.

In this study, we propose methods – based on Cited References Analysis (CRA) and Reference Publication Year Spectroscopy (RPYS) – to identify those publications in a research field or on a specific topic which have been influential over many years in the past. Thus, the methods – which have been implemented in the bibliometric tool CitedReferencesExplorer (CRExplorer at http://www.crexplorer.net) – identify those publications (papers, books, reports etc.), which were highly cited over a longer time period or at certain time points (shortly or several years after publication). In these analyses, different types of citation distributions are considered to identify, e.g., publications receiving many citations very rapidly ("hot papers"), several years after appearance ("sleeping beauties"), or across the whole life span ("constant performers"). With information on these types, the user of the CRExplorer receives additional information on a paper's impact, which are beyond the usual citation impact (or cited references) analysis.

Similar methods of identifying landmark papers in a set of papers have been published by Mazloumian, Eom, Helbing, Lozano, and Fortunato (2011) and Bornmann, Ye, and Ye (in press). However, these methods focus on the times cited and not the CRs perspective.

# 2 Cited references analysis (CRA) and Reference Publication Year Spectroscopy (RPYS)

The starting point of CRA is the selection of the publication set representing a specific field (e.g. bibliometrics) or dealing with a specific topic (e.g. research on Aspirin). Then, the CRs are extracted and their occurrences are cumulated. The most important advantage of CRA against the times cited analysis is the target-oriented impact measurement: The bibliometrician defines the target on which impact is intended to be measured by selecting the publications of a field or topic. The more competently this publication set is compiled, the more the bibliometrician is able to identify the influential publications in a field or topic (measured in terms of CR counts) (see the explanations of Haunschild, Bornmann, & Marx, 2016).

The CRA approach agrees with the usual aim of many citation impact studies, which are intended to measure the impact of contributions on a specific field. For example, "the result is the identification of high performers within a given scientific field" (Froghi et al., 2012, p. 321). "Ideally, a measure would reflect an individual's relative contribution within his or her field" (Kreiman & Maunsell, 2011). "That is, an account of the number of citations received by a scholar in articles published by his or her field colleagues" (Di Vaio, Waldenström, & Weisdorf, 2012, p. 92).

In contrast to the CRA, which measures the impact on a well-defined target (field or topic), the times cited analysis measures the impact in the whole scientific community. The focus on a well-defined target obviates the normalization of citation impact which is necessary in the times cited analysis: professional bibliometrics without normalization is difficult to imagine because impact using times cited data is mostly measured across different fields. The focus on one field (or topic) in the CRA implies field normalization and avoids advanced methods of field normalization, which are described by Waltman (2016) or Bornmann and Haunschild (2016).

A specific kind of CRA is RPYS, which is explained by Marx, Bornmann, Barth, and Leydesdorff (2014) as follows: "RPYS is based on the analysis of the frequency with which references are cited in the publications of a specific research field in terms of the publication years of these CRs. The origins show up in the form of more or less pronounced peaks mostly caused by individual publications that are cited particularly frequently" (p. 751). The CRExplorer was developed by Thor et al. (2016a, 2016b) for CRAs and RPYS using Web of Science (WoS, Clarivate Analytics) or Scopus (Elsevier) data.

In recent years, RPYS has been applied in a variety of contexts: Wray and Bornmann (2014) investigate the roots of the philosophy of science, Marx et al. (2014) the origins of graphene and solar cell research, Rhaiem and Bornmann (2018) citation classics in the area of academic efficiency studies, Ballandonne (2018) the historical roots of contributions to ecological economics, and Elango, Bornmann, and Kannan (2016) the seminal publications in modern tribology research. Furthermore, Marx and Bornmann (2014) used RPYS to demask a scientific legend, and Comins and Hussey (2015a, 2015b) analyzed the landmark research contributions to the global positioning system (GPS) and investigated the impact of the Viterbi algorithm.

Most RPYS publications published hitherto have focused on the history in a scientific field or topic (on the 19th and the first half of the 20th century). In the era of little science (before around 1950, see Marx & Bornmann, 2010) the number of CRs in a field or topic is comparatively low, which facilitates the identification of important contributions. However, in the big science period, the growth of literature leads to numerous CRs whereby the important contributions are difficult to identify by RPYS. For purposes of analyzing the complete range of contributions, Comins and Leydesdorff (2016) introduced the Multi-RPYS, which segments "the set of citing articles by their publication years and performing a standard RPYS analysis for each year under study. The results are then rank-transformed and organized in a heatmap to visualize the dynamic influences of cited references on the citing set" (p. 1511).

Multi-RPYS is a major step in RPYS development, which allows the investigation of communal intellectual histories and temporal dynamics of historical influences. The heat maps provided by RPYS i/o enable a comprehensive overview on the most important RPYs for the citing years under study. However, RPYS i/o can scarcely be used to identify the single most important publications in a field or for a topic. Thus, we extended the CRExplorer with an advanced statistics segment which operates on the single publication level. Applying various advanced statistics to the dataset, the user of the CRExplorer is able to identify the most influential single publications (e.g. hot paper, constant performers, or sleeping beauties) over different bands of citing publication years.

# 3 Methods

## 3.1 CitedReferencesExplorer (CRExplorer)

The CRExplorer was specifically developed by Thor et al. (2016a, 2016b) for analyzing the CRs in a specific publication set (downloaded from Scopus or WoS). In recent years, two other programs have been introduced enabling CR analyses: RPYS i/o (see http://comins.leydesdorff.net) (Comins & Leydesdorff, 2016) and metaknowledge (see http://networkslab.org/metaknowledge) (McLevey & McIlroy-Young, 2017). Datasets from WoS or Scopus (publications including CRs) can be uploaded in the CRExplorer. The program visualizes the number of CRs per reference publication year and tabs the CRs. The user of the program can select single reference publication years (RPYs) in the visualization and the corresponding CRs are highlighted in the table. Thus, the user is able to identify the publications behind RPYs producing more citation impact than other years.

The functionality of the CRExplorer is adjusted to the practice of CRA. Thus, the user can utilize the program to prepare the dataset for the statistical analysis: For example, the dataset can be limited to the CRs with larger impact and the CRs can be disambiguated. The possibility of disambiguation is a specific feature of the CRExplorer, which allows the clearing of the dataset from variants of the same CR. The existence of variants in the data is a major problem in citation analysis, which might lead, e.g., to an underestimation of the impact of books. Books are typical document types affected by many variants. Several publications (e.g., Moed, 2005; Olensky, Schmidt, & van Eck, 2016) in bibliometrics have pointed to the problem with CR data that there exist variants of the same CR. For example, Moed (2005) investigated 22 million CRs from the WoS and found 7.7% discrepant CRs resulting in a missed match with target papers. The disambiguation is especially necessary for Scopus data; however, the Scopus data is especially suitable for disambiguation, because the title of the CR can be considered (which is not possible with WoS data). This allows the disambiguation on a broader data basis.

The CRExplorer supports the analysis of the CRs by sorting the CRs of a specific RPY by citation impact in decreasing order. This allows rapid identification of the most important CRs. Furthermore, the CR data is visualized in a way that can be adapted to the need of the user. Not only the data for the visualization can be downloaded for processing with other programs, the (revised) CR data can be saved in a specific file format of the CRExplorer or in the WoS or Scopus format. Thus, it is possible to upload Scopus data to the CRExplorer, process the data in the program, and download it in the WoS data format.

### 3.2 Example dataset

At the beginning of 2017, we downloaded from Scopus 5506 papers (including CR data), which were published in *Scientometrics* between 1978 and 2016. We considered all document types. We decided to use this publication set as an example in this study, since *Scientometrics* is the oldest journal dedicated to the field of scientometrics (starting in 1978). We are interested in the impact of specific CRs over several publication years. Obviously, this analysis is restricted by the publication years of the citing publications. The long publication history of *Scientometrics* allows the analysis of the impact of CRs over a long period. Other journals in the field of scientometrics (e.g., *Journal of Informetrics*) offer only significantly shorter time periods.

Before we started to analyze the data using the new functionalities, we revised the dataset in several steps (which are normally necessary for CRA). Since this study focusses on temporal dynamics of historical influences, we selected the uploaded range of CRs from 1900 to 2005 (resulting in n=66,617 CRs). In a first step, we cleared the dataset of variants of the same CR using the matching and clustering facilities by the CRExplorer. These facilities are explained in detail by Thor et al. (2016a).

Two CRs are considered to match, if their similarity is above a user-defined threshold (e.g., 75%). To this end, CRExplorer computes the pair-wise string similarities of title (if available), authors' last names, and source title. The similarity values are aggregated then to an overall similarity value. The combination of multiple similarity values that are based on different attributes typically achieves a better match quality compared to a single similarity of the entire CR strings (Köpcke, Thor, & Rahm, 2010). Finally, CRExplorer performs a clustering based on the matching results, i.e., the list of the matching CR pairs. Two CRs are assigned to the same cluster if they are matching or if they are both matching other CRs that are already assigned to the same cluster. During the data cleaning process, only one representative remains in the dataset for each cluster. From the variants (CRs) forming a cluster the one variant is selected as representative which has the highest number of occurrences in the cluster. The numbers of occurrences for all variants of the cluster are summarized and assigned to the representative.

The matching and clustering process reduced the dataset of this study to n=44,123 CRs. In a second step, we deleted all CRs for which the bibliographic information did not match the categorization used by the CRExplorer (i.e. authors, publication year, title etc.). These CRs can be identified by sorting the CR data by the authors and deleting the CRs without authors or with obviously wrong author information (e.g. the author field contains a title fragment). Furthermore, some variants of the same CRs have been manually aggregated. The second step leads to the final dataset of n=33,812 CRs.

# 4 Results

In the following, we present the new advanced statistics in the CRExplorer. It is the general objective of the statistics to identify influential papers in the publication set. The impact of the papers is measured across the publication period of the citing papers. The new statistics have been included in the program by adding columns to the table on the right side of the screen. Using the menu item “File” – “Settings” – “Table” (section “Indicators”), the columns can be visualized or suppressed. The CRExplorer newly computes all indicator values if any changes are made to the dataset (e.g. if CRs are deleted or clustered).

### 4.1 Top 50%, 25%, and 10% cited references in citing years

The CRExplorer has been initially programmed to identify the most influential RPYs (the peak years) and the CRs (cited publications) which essentially produced the peaks in these years. Here, the impact of the cited publications is measured across all citing publications in the dataset. Since most of the impact is generated in the first three to five years after publication, the influential publications are frequently important in the field for only a few years after publication. Thus, it is additionally interesting to identify those exceptional publications (top publications), which are important (influential) over a longer time period. The functionality of the CRExplorer has been extended to facilitate this objective.

Table 1. Small-world example for explaining top 50%, 25%, and 10% cited references in citing years

| | Publication year of citing publication | | | | | | Results |
|---|---|---|---|---|---|---|---|
| RPY=1970 | 1980 | 1981 | 1982 | 1983 | 1984 | 1985 | N_PYEARS |
| CR A | 6 | 5 | 0 | 17 | 24 | 21 | 5 |
| CR B | 9 | 9 | 5 | 10 | 8 | 9 | 6 |
| CR C | 20 | 34 | 0 | 16 | 5 | 6 | 5 |
| CR D | 6 | 10 | 15 | 25 | 15 | 5 | 6 |
| | Sorted cells in ascending order (rows) | | | | | | |
| | 6 | 5 | 0 | 10 | 5 | 5 | |
| | 6 | 9 | 0 | 16 | 8 | 6 | |
| | 9 | 10 | 5 | 17 | 15 | 9 | |
| | 20 | 34 | 15 | 25 | 24 | 21 | |
| | Limits for identifying top publications | | | | | | |
| top 50% | 6 | 9 | 0 | 16 | 8 | 6 | |
| top 75% | 9 | 10 | 5 | 17 | 15 | 9 | |
| top 90% | 9 | 10 | 5 | 17 | 15 | 9 | |
| | Values below and above the limit (top 50%) | | | | | | N_TOP50 |
| CR A | 0 | 0 | 0 | 1 | 1 | 1 | 3 |
| CR B | 1 | 0 | 1 | 0 | 0 | 1 | 3 |
| CR C | 1 | 1 | 0 | 0 | 0 | 0 | 2 |
| CR D | 0 | 1 | 1 | 1 | 1 | 0 | 4 |
| | Values below and above the limit (top 25%) | | | | | | N_TOP25 |
| CR A | 0 | 0 | 0 | 0 | 1 | 1 | 2 |
| CR B | 0 | 0 | 0 | 0 | 0 | 0 | 0 |
| CR C | 1 | 1 | 0 | 0 | 0 | 0 | 2 |
| CR D | 0 | 0 | 1 | 1 | 0 | 0 | 2 |
| | Values below and above the limit (top 10%) | | | | | | N_TOP10 |
| CR A | 0 | 0 | 0 | 0 | 1 | 1 | 2 |
| CR B | 0 | 0 | 0 | 0 | 0 | 0 | 0 |
| CR C | 1 | 1 | 0 | 0 | 0 | 0 | 2 |
| CR D | 0 | 0 | 1 | 1 | 0 | 0 | 2 |

We start by explaining the methods for identifying the time period of influence by using the small world example in Table 1. The small world consists of four CRs (A, B, C, and D), which have been published in 1980 and cited in 1980, 1981, 1982, 1983, 1984, and 1985. For example, CR A has been cited in 5 publications, which were published in 1981. The first new indicator in the CRExplorer, named N_PYEARS, is equal to the number of years in which a CR has been cited. In the small world, the CR A has been cited in five citing years. Thus, N_PYEARS=5 for CR A. The user of the CRExplorer should be aware that the number of citing years is defined by the publication years of the citing publications. For example, a CR from 1990 can only be cited in 10 years (and not 20 years), if the underlying dataset includes publications from 2000 to 2009. In order to call the attention of the CRExplorer user to these limitations defined by the range of publication years in the dataset, the status bar shows not only the range of the RPYs, but also the range of the publication years of the citing publications (maximal number of citing years). The second new indicator in the CRExplorer – named PERC_PYEAR – is the percentage of years in which the CR has been cited. Thus, N_PYEARS is divided by the maximal number of citing years (i.e., all publication years with at least one citation to a CR in RPY) to yield PERC_PYEAR (not shown in Table 1).

PERC_PYEAR highlights those CRs which received at least one citation in many citing years. However, we are further interested in those CRs which have been cited more frequently in the citing years than other CRs in the dataset. In order to identify these CRs, thresholds are computed which identify the top 50%, top 25%, and top 10% in one citing year. In the first step of the computation, the citations in one citing year are sorted in ascending order (see Table 1). In the second step, the thresholds for the top 50%, 25%, and 10% are determined in a given year. In the third step, those CRs are identified which are above the three thresholds. In the fourth step, the numbers of citing years are counted in which the CRs are above the thresholds. These numbers yield N_TOP50, N_TOP25, and N_TOP10.

It might be a problem in computing N_TOP50, N_TOP25, and N_TOP10 if the citation counts in a citing year are inflated by zeros (and/or similar values). Thus, we included the option in the CRExplorer to extend the number of citing years which are considered in calculating N_TOP50, N_TOP25, and N_TOP10. The number of citing years can be set in the menu item “File” – “Settings” – “Table” – “NPCT Range” in section “Value settings”. If only the citing year itself should be considered in the analysis, the “NPCT Range” is set to 0 (as done in Table 1). If it is set to 1, the thresholds for the top 50%, 25%, and 10% are computed on the basis of the citations from the preceding ($t$-1) and succeeding ($t$+1) citing years. This doubles the underlying dataset in the first and last citing year (since year $t$-1 and $t$+1, respectively, are considered) and triples it in the years in-between.

In the *Scientometrics* dataset, the Lotka (1926) paper on the distribution of scientific productivity and the de Solla Price (1963) book “little science, big science” are those publications with the highest number of years in which they have been cited by other publications (N_PYEARS=36). Both publications appear at the top of the table in the CRExplorer if the CRs are sorted by the column N_PYEARS. It follows Garfield (1979) with N_PYEARS=34 at the third position. However, the percentages in the column PERC_PYEAR point out that they have not been cited in all possible years (39 years: 1978-2016, see the corresponding information in the status bar). If we sort the CRs by the column

PERC_PYEAR, we identify 13 publications with PERC_PYEAR=100%, which are listed in Table 2.

Table 2. Thirteen publications in the field of scientometrics with at least one citation in every year since their publication (identified by the references cited in *Scientometrics* papers)

| Cited reference | Title | Publication medium | N_CR | N_ PYEARS |
|---|---|---|---|---|
| Hirsch (2005) | An index to quantify an individual's scientific research output | *PNAS* | 403 | 12 |
| Katz and Martin (1997) | What is research collaboration? | *Research Policy* | 171 | 20 |
| Lotka (1926) | The frequency distribution of scientific productivity | *Journal of the Washington Academy of Sciences* | 155 | 36 |
| Moed (2005) | Citation analysis in research evaluation | Book | 137 | 12 |
| Schubert and Braun (1986) | Relative indicators and relational charts for comparative assessment of publication output and citation impact | *Scientometrics* | 120 | 31 |
| van Raan (2005) | Fatal attraction: conceptual and methodological problems in the ranking of universities by bibliometric methods | *Scientometrics* | 82 | 12 |
| Glänzel and Schubert (2003) | A new classification scheme of science fields and subfields designed for scientometric evaluation purposes | *Scientometrics* | 66 | 14 |
| Persson, Glänzel, and Danell (2004) | Inflationary bibliometric values: the role of scientific collaboration and the need for relative indicators in evaluative studies | *Scientometrics* | 66 | 13 |
| Glänzel and Schubert (2001) | Double effort = double impact? A critical view at international co-authorship in chemistry | *Scientometrics* | 57 | 16 |
| Egghe (2005) | Power laws in the information production process: Lotkaian informetrics | Book | 46 | 12 |
| Glänzel and Schubert (2004) | Analyzing scientific networks through co-authorship | Handbook of Quantitative S & T Research. | 43 | 13 |
| Weingart (2005) | Impact of bibliometrics upon the science system: inadvertent consequences? | *Scientometrics* | 35 | 12 |
| Jin and Rousseau (2005) | China's quantitative expansion phase: Exponential growth, but low impact | 10th ISSI conference | 18 | 12 |

Notes. N_CR=Number of occurrences, N_PYEARS=Number of years in which the publication has been cited

Table 2 shows some important publications in the field of scientometrics, which deal – among other things – with collaboration in research, university rankings, normalized indicators, and the role of China in the worldwide science system. Also, the paper introducing the *h* index is among these papers. The 13 publications have not only been published in journals (*Scientometrics* and *Research Policy*), but also as books, in a handbook, and in the proceedings of the 10th ISSI conference.

Table 3. Ten publications in the field of scientometrics with the highest number of citing years in which they belong to the 10% most frequently cited publications (identified by the references cited in *Scientometrics* papers)

| Cited reference | Title | Publication medium | N_CR | N_TOP10 | PERC_ PYEARS |
|---|---|---|---|---|---|
| Lotka (1926) | The frequency distribution of scientific productivity | *Journal of the Washington Academy of Sciences* | 155 | 36 | 100.00 |
| de Solla Price (1963) | Little science, big science | Book | 216 | 36 | 94.74 |
| Garfield (1979) | Citation indexing: its theory and application in science, technology, and humanities | Book | 151 | 34 | 91.89 |
| Small (1973) | Co-citation in the scientific literature: a new measure of the relationship between two documents | *Journal of the American Society for Information Science* | 162 | 33 | 84.62 |
| Cole and Cole (1973) | Social stratification in science | Book | 74 | 32 | 82.05 |
| Schubert and Braun (1986) | Relative indicators and relational charts for comparative assessment of publication output and citation impact | *Scientometrics* | 120 | 31 | 100.00 |
| Garfield (1972) | Citation analysis as a tool in journal evaluation: journals can be ranked by frequency and impact of citations for science policy studies | *Science* | 109 | 31 | 79.49 |
| Small and Griffith (1974) | The Structure of Scientific Literatures I: Identifying and Graphing Specialties | *Science Studies* | 69 | 31 | 79.49 |
| Narin (1976) | Evaluative bibliometrics: the use of publication and citation analysis in the evaluation of scientific activity | Book | 48 | 30 | 76.92 |
| Merton (1968) | The Matthew effect in science | *Science* | 115 | 30 | 76.92 |

Notes. N_CR=Number of occurrences, N_TOP10= Number of citing years in which they belong to the 10% most frequently cited publications, PERC_PYEAR=Percentage of years in which the publication has been cited. The “NPCT Range” is set to 0 in the CRExplorer.

Table 3 shows the ten publications in the field of scientometrics with the highest number of citing years in which they belong to the 10% most frequently cited publications. There is only one publication in Table 3 which is also in Table 2: The paper by Schubert and Braun (1986) about the introduction of field-normalization in bibliometrics. Table 3 lists not only publications which are groundbreaking in bibliometrics, such as the paper by Schubert and Braun (1986), the paper by Small (1973) about the introduction of the method of co-citation, and the first published journal ranking on the basis of the JIF (Garfield, 1972); it lists also classics from the sociology of science. These include the introduction of the Matthew effect (Merton, 1968) and the explanation of the consequences which result from the social stratification system in the scientific community (Cole & Cole, 1973).

### 4.2 Sequence analysis

Besides the question of identifying exceptionally influential publications (top publications) it is also of interest to identify the citation dynamic of CRs (Bornmann, Ye, & Ye, 2017). Usually, cited publications have a lifetime with the following dynamic: starting with low citations in the first year of publication, growing up to a maximum of citations a few years later, followed by a continuous decrease of citations several years after publication (Redner, 1998). However, other dynamics are also possible: a more or less long period of non-recognition with low citations is followed by a period with high citations after a sudden peak. Such a dynamic is typical for the phenomenon named “sleeping beauty” (van Raan, 2004), “for publications whose importance is not recognized for several years after publication”(Ke, Ferrara, Radicchi, & Flammini, 2015b, p. 7426).

In order to identify statistically the citation dynamics of CRs with the CRExplorer, we apply Configural Frequency Analysis (CFA, Stemmler, 2014; von Eye, 2002; von Eye, Mair, & Mun, 2010). CFA is a categorically statistical procedure to reveal configurations in multivariate cross-classifications (i.e., contingency tables). The CRs for a certain RPY and the publication year for the citing publications are cross-classified, as shown in our small-world example (see Table 4), with the citation count for each combination in the cell. CFA focusses on the individual cells of a contingency table instead of the variables (rows, columns) establishing the table.

In the case of systematic citation dynamics (e.g., lifetime cycles) citations in the cells deviate strongly from the expected values. Expected frequencies are cell frequencies which would occur if there is no relationship between or independency of the row (CRs) and the column variable (publication year). These expected frequencies can be calculated by multiplying the marginal frequencies for the corresponding row and column of each cell, and by further dividing the product by the overall frequency (see Table 4, Expected). For instance, in order to obtain the expected value of 15.12 for the cell “publication year 1981” and “cited references A” the corresponding row frequency of 73 is multiplied by the column frequency of 58. The resulting product is further divided by the total frequency of 280 (=73*58/280=15.12).

Expected values should usually be greater than 5. As a measure of deviance from the independency-base model the Pearson-$\chi^2$ is used. The Pearson-$\chi^2$ is defined as the sum of the squared deviances of the observed (o) from the expected values (e) of each cell, divided by the expected value: $\chi^2$ (df=(r-1)(c-1)= $\Sigma$(o-e)$^2$/e, where r is the number of rows and c is the number of columns in the contingency tables. In order to characterize a specific cell, z-values are calculated: z= (o-e)/√(e), where $\chi^2 = \Sigma z^2$. For example, for the first cell (see Table 4, z-value) the z-value of -1.43 is obtained by dividing the difference between observed (=6) and expected (=10.69) value by 10.69 (=(6-10.69)/√10.69=-1.43). Actually, z-values are standard normally distributed with mean value of zero and standard deviation of 1.0. High positive or negative z values identify cells which strongly deviate from the independency-base model, and they indicate a certain citation dynamic: “types” with positive z-values and “antitypes” with negative z-values in the terminology of CFA by von Eye et al. (2010).

In our case, the absolute z-value of 1.0 (one standard deviation) provides a threshold to identify cells with significant deviations. Other thresholds are possible as well, for example, a z-value of 1.96 (5% probability that the deviation occurs under the condition of independency of rows and columns). Statistical inference is used here solely for pattern recognition to reveal signals in the noise, not to make any inference about a population of interest.

Table 4. Small-world example for explaining the rationale of configuration frequency analysis (CFA)

| Observed | Publication year of citing publication | | | | | | Row frequency |
|---|---|---|---|---|---|---|---|
| RPY=1970 | 1980 | 1981 | 1982 | 1983 | 1984 | 1985 | |
| CR A | 6 | 5 | 0 | 17 | 24 | 21 | 73 |
| CR B | 9 | 9 | 5 | 10 | 8 | 9 | 50 |
| CR C | 20 | 34 | 0 | 16 | 5 | 6 | 81 |
| CR D | 6 | 10 | 15 | 25 | 15 | 5 | 76 |
| Column frequency | 41 | 58 | 20 | 68 | 52 | 41 | 280 |
| | | | | | | | |
| Expected | Publication year of citing publication | | | | | | Row frequency |
| RPY=1970 | 1980 | 1981 | 1982 | 1983 | 1984 | 1985 | |
| CR A | 10.69 | 15.12 | 5.21 | 17.73 | 13.56 | 10.69 | 73 |
| CR B | 7.32 | 10.36 | 3.36 | 12.14 | 9.29 | 7.32 | 50 |
| CR C | 11.86 | 16.78 | 5.78 | 19.67 | 15.04 | 11.86 | 81 |
| CR D | 11.13 | 15.74 | 5.43 | 18.46 | 14.11 | 11.13 | 76 |
| Column frequency | 41 | 58 | 20 | 68 | 52 | 41 | 280 |
| | | | | | | | |
| z-value | Publication year of citing publication | | | | | | |
| RPY=1970 | 1980 | 1981 | 1982 | 1983 | 1984 | 1985 | |
| CR A | -1.43 | -2.60 | -2.28 | -0.17 | 2.84 | 3.15 | |
| CR B | 0.62 | -0.42 | 0.76 | -0.61 | -0.42 | 0.62 | |
| CR C | 2.36 | 4.20 | -2.41 | -0.83 | -2.59 | -1.70 | |
| CR D | -1.54 | -1.45 | 4.11 | 1.52 | 0.24 | -1.84 | |
| | | | | | | | |
| Sequence | Publication year of citing publication | | | | | | |
| RPY=1970 | 1980 | 1981 | 1982 | 1983 | 1984 | 1985 | |
| CR A | - | - | - | 0 | + | + | Type 1 |
| CR B | 0 | 0 | 0 | 0 | 0 | 0 | Type 2 |
| CR C | + | + | - | 0 | - | - | Type 3 |
| CR D | - | - | + | + | 0 | - | Type 1 |

Note: “+”=z>1, “-“=z<-1, otherwise 0.

In order to reveal specific sequences over time, rows of cells (CR) are considered with average (“0”; -1≤z≤1), above average (“+”; z>1), and below average (“-”; z<-1) cells, whereby average is used here in the sense of expected values. Based on the sequences, types of CRs in terms of different citation dynamics or sequences of symbols (“+”, ”-“, “0”) can be identified (see Table 4, Sequence), which are labelled as follows: “sleeping beauty” with low or no citations over a longer initial period and high citations later (type 1), “constant performer” with a constant and considerable amount of citations over time (type 2), “hot paper” with high citations directly after the publication and low citations later (type 3), and “life cycle” with courses of different annual citations across time (type 4). If CRs belong to more than one type, all types are indicated in the table of the CRExplorer.

Table 5. Definition and default parameters for identifying different types of sequences and example publications from scientometrics, which belong to the types (using these parameters)

| | |
|---|---|
| **Sleeping beauty (type 1)** | Publication which has been cited below average in two of the first three citing years (“-“; z<-1) and above average (“+”; z>1) in the following citing years at least once |
| Barabasi et al. (2002) | Title: Evolution of the social network of scientific collaborations<br>Sequence: 0--0---0++0-0++ |
| Girvan and Newman (2002) | Title: Community structure in social and biological networks<br>Sequence: 0----00000000+0 |
| **Constant performer (type 2)** | Publication which has been cited in more than 80% of the citing years at least once. In more than 80% of the citing years it has been cited at least on the average level (“0”; -1<=z<=1) or (“+”; z>1) |
| Lotka (1926) | Title: The frequency distribution of scientific productivity<br>Sequence: 000000000000000000000000000-0000000000 |
| Moed (2005) | Title: Citation analysis in research evaluation<br>Sequence: 0-000000+000 |
| **Hot paper (type 3)** | Publication which has been cited above average (“+”; z>1) in two of the first three citing years after publication |
| Bornmann and Daniel (2005) | Title: Does the h-index for ranking of scientists really work?<br>Sequence: 0++++0000--- |
| Braun, Glänzel, and Schubert (2005) | Title: A Hirsch-type index for journals<br>Sequence: 0+++000+---0 |
| **Life cycle (type 4)** | Publication which has been cited in at least two of the first four years on the average level (“0”; -1<=z<=1) or lower (“-”; z<-1), in at least two years of the following years above average (“+”; z>1), and in the last three years on the average level (“0”; -1<=z<=1) or lower (“-”; z<-1) |
| Hirsch (2005) | Title: An index to quantify an individual’s scientific research output<br>Sequence: -0--+++++0-0 |
| de Solla Price (1963) | Title: Little science, big science<br>Sequence: 0+00--0000-0000++00+00++000000000+--000 |

The detailed definitions of the different types of sequences are presented in Table 5. For example, “hot papers” are those which have been cited above average in the first three years after publication. Table 5 shows not only the definitions of the types, but presents also some type examples in the *Scientometrics* dataset. For example, the paper by Barabasi et al. (2002) has been cited above average several years after appearance. The first years are characterized by below average citations. This type of citation distribution is called “sleeping beauty” in scientometrics (van Raan, 2004).

Several publications in the past have targeted this citation impact type. Authors have been fascinated by the fact that publications remained undetected over many years, before the results, methods, ideas etc. become important for current research. A couple of case studies have been published describing certain cases of sleeping beauties (e.g., Gorry & Ragouet, 2016; Marx, 2014; Tal & Gordon, 2017). Ke, Ferrara, Radicchi, and Flammini (2015a) and Ye and Bornmann (2018) have published variants of definitions of how sleeping beauties can be identified in publication sets (see also Goldstein, 2017). In a recent study, van Raan (2015) found that many sleeping beauties are application-oriented, which means that they are potential sleeping innovations. In a follow-up study, van Raan (2016) analyzed characteristics of sleeping beauties which have been cited in patents.

The use of the “hot papers” concept is especially connected to the WoS database. For every publication set which has been selected in the WoS database, hot papers are marked with a symbol and counted. Clarivate Analytics defines hot papers as “papers published in the past two years that are in the top one-tenth of one percent (0.1%) for their field and publication period” (see https://clarivate.com/blog/new-hot-papers-may-2017). These papers have a very early citation peak and later annual citation rates which are significantly lower than the early peak (Ye & Bornmann, 2018). In the *Scientometrics* dataset, Braun et al. (2005) was assigned to the “hot paper” type, since the paper had an early peak and low(er) later citation rates (compared to publications from the same year). Several reasons can lead to the decrease of citations after the initial high-impact phase: (1) The interest of the community in the topic of the paper declines. (2) The results of the paper could not be replicated in other studies. (3) The paper is concerned by the “obliteration by incorporation” phenomenon (Garfield, 1975; McCain, 2011, 2015), whereby certain ideas are incorporated into the accepted archive of knowledge and no longer cited.

Table 5 includes two further types, which are diametrically opposed. “Constant performers” are characterized by citation rates which are constantly at least on the average level – compared to the other publications. Publications of the “Life cycle” type start with relatively low citation rates, have a relatively high impact later on and finish with relatively low citation rates. The paper by Hirsch (2005) shows these characteristics, as the sequence in Table 5 reveals.

# 5 Discussion

In RPYS, CRs of publication sets are analysed to identify the most important contributions in the past. Alternative concepts to RPYS for analysing historical papers have been proposed since the 1990s. Most important are the concepts of co-citations (Small & Griffith, 1974) and research fronts (de Solla Price, 1965) as well as the method named “algorithmic historiography” (Garfield, Pudovkin, & Istomin, 2003; Leydesdorff, 2010). The HistCite™ software (Garfield, 2009), which has been developed by Alexander Pudovkin and Eugene Garfield for “algorithmic historiography”, visualizes the citation network among publication sets, including historical papers. With the CitNetExplorer, a program similar to HistCite™ has been developed by van Eck and Waltman (2014). It analyses and visualizes citation networks of a given publication set (see www.citnetexplorer.nl). RPYS with CRExplorer focusses on the citation impact distribution of single publications, but does not compute networks of CRs. RPYS reveals quantitatively which historical papers are of particular importance for a given publication set.

The proposal to perform impact analyses from the CRs rather than the “times cited” view is based on the idea that the analysis should focus on the impact one gets from direct peers. These are researchers working and publishing on the same topic or similar topics. The analysis of CRs for impact measurements is not a new approach in bibliometrics, but can already be found in de Solla Price (1963). Other studies have used the CR approach to answer specific research questions, for example to measure growth rates of science (Bornmann & Mutz, 2015; van Raan, 2000). Growth rates should actually be calculated on the basis of publication numbers. However, these numbers are only available for the past decades. The switch to CRs for measuring growth rates means that (referenced) publications from (very) early years can be considered in the analysis. The disadvantage of the approach is that only cited publications can be considered.

RPYS has been developed for identifying the CRs with the greatest influence in a given paper set (mostly sets of papers in certain topics or fields). With the former versions of the CRExplorer, the search for these CRs was dependent on the visual inspection of the spectrogram provided by the program. The user had to inspect the CRs underlying the peaks in order to select the most influential publications. In early RPYs, peaks are mostly triggered by the impact of single CRs. In other words, influential CRs can be properly identified in these years by visual inspection. However, in more recent RPYs, many CRs contribute to single peaks to a similar extent, which make it difficult to select single influential CRs. As a possible solution for the problem of identifying the influential CRs (especially in recent years), Comins, Carmack, and Leydesdorff (2017) proposed calculating an indicator for every CR in the set whereby the proportion of occurrences of a CR in the corresponding RPY is weighted by the median deviation. This is the deviation of the number of CRs in the focal year (Y) from the median for the number of CRs in the X previous, the current, and the X following years. However, the weighted indicator proposed by Comins et al. (2017) refers to the CRs counts in total and does not consider the influence of CRs over the series of citing publication years.

In this study, we have presented some methods to identify and characterize CRs which have been influential across a longer period (several citing years). The indicators N_TOP50, N_TOP25, and N_TOP10 proposed in CRExplorer can be used to inspect those CRs with (significantly) higher impact than comparable CRs from the same RPY. Indicator values of more than 10 or 20 reveal CRs which belonged to the most highly cited over 10 or 20 citing publication years. The analysis of the example dataset revealed, for example, that the paper by Lotka (1926) entitled "The frequency distribution of scientific productivity" belongs to the 10% most frequently cited publications in 36 citing years. Thus, this paper seems to be of general importance for the field of scientometrics. However, papers such as Lotka (1926), are exceptions; many publications show citation distributions which are characterized by changes in citation impact intensities over the citing years. Therefore, the new version of CRExplorer analyses the sequence of citations across the given citing years to identify different types. The sequence is used by the program to identify papers with typical impact distributions. For example, publications can have early, but not late impact (hot papers) or vice versa (sleeping beauties).

The impact analysis of historical papers has two limitations which should be considered in applying RPYS with CRExplorer (McCain, 2011, 2015): "obliteration by incorporation" and "palimpsestic syndrome". Both phenomena go back to Merton (1965). The first phenomenon describes a process by which results, ideas, or methods from seminal publications have been (quickly) absorbed into the body of knowledge in a field or on a topic. The content from these publications has been heavily used not only in research papers, but also in textbooks without citing the original source. The content has become basic knowledge. The second phenomenon describes a process by which it is no longer the initial publications of results, ideas, or methods which are cited, but later publications, which cite these initial publications. Both phenomena might lead to a reduction of citation impact for landmark papers, which should be considered in the interpretation of RPYS results. However, the reduction of impact through the influence of these phenomena is not so large that the general evidence of the results has to be questioned.

# 6 Conclusions

We explained some advanced methods which have been newly developed for CRExplorer. These methods identify and characterize the CRs which have been influential across many citing years. The indicators N_TOP50, N_TOP25, and N_TOP10 can be used to identify those CRs which belong to the 50%, 25%, or 10% most frequently cited publications over many citing publication years. In the *Scientometrics* dataset, for example, Lotka's (1926) paper on the distribution of scientific productivity belongs to the top 10% publications in 36 citing years. Furthermore, the new version of CRExplorer analyzes the impact sequence of CRs across citing years. CRs can have below average (-), average (0), or above average (+) impact in citing years (whereby average is meant in the sense of expected values). The sequence (e.g. 00++---0--00) is used by the program to identify publications with typical impact distributions. For example, CRs can have early, but not late impact ("hot papers", e.g. +++---) or vice versa ("sleeping beauties", e.g. ---0000---++).